\documentclass{article}
\usepackage{graphicx}

\def\giorno{2 April 2006}

\def\a{\alpha}

\def\ga{\gamma}
\def\de{\delta}   

\def\la{\lambda}
\def\s{\sigma}

\def\om{\omega}
\def\th{\theta}
\def\vth{\vartheta}
\def\vphi{\varphi}

\def\pa{\partial}
\def\d{{\rm d}}       

\def\<{\langle}
\def\>{\rangle}

\def\({\left(}
\def\){\right)}
\def\[{\left[}
\def\]{\right]}
\def\=#1{\bar #1}
\def\~#1{\widetilde #1}
\def\.#1{\dot #1}
\def\^#1{\widehat #1}
\def\"#1{\ddot #1}

\begin{document}

\title{Solitons in Yakushevich-like models of DNA dynamics
with improved intrapair potential\thanks{Work supported in part by
the Italian MIUR under the program COFIN2004, as part of the PRIN
project {\it ``Mathematical Models for DNA Dynamics ($M^2 \times
D^2$)''}.}}

\author{Giuseppe Gaeta \\
{\it Dipartimento di Matematica, Universit\`a di Milano} \\
{\it via Saldini 50, 20133 Milano (Italy)} \\
gaeta@mat.unimi.it}

\date{\giorno}

\maketitle
\def\Y{Yakushevich }

\noindent {\bf Summary.} The \Y model provides a very simple
pictures of DNA torsion dynamics, yet yields remarkably correct
predictions on certain physical characteristics of the dynamics.
In the standard \Y model, the interaction between bases of a pair
is modelled by a harmonic potential, which becomes anharmonic when
described in terms of the rotation angles; here we substitute to
this different types of improved potentials, providing a more
physical description of the H-bond mediated interactions between
the bases. We focus in particular on soliton solutions; the \Y
model predicts the correct size of the nonlinear excitations
supposed to model the ``transcription bubbles'', and this is
essentially unchanged with the improved potential. Other features
of soliton dynamics, in particular curvature of soliton field
configurations and the Peierls-Nabarro barrier, are instead
significantly changed.

\section*{Introduction}

In the DNA transcription process, a specialized enzyme
(RNA-Polymerase) binds to a specific site of the DNA double helix
and unwinds it in a local region of 15-20 bases, thus creating a
``transcription bubble''; the RNAP and the bubble travel then
along the DNA, copying its sequence and producing a RNA-Messenger
to be later used to express genes or replicate the local sequence.

This process requires a very finely tuned coordination of the
motion of RNAP -- and production of the RNA-Messenger -- with the
dynamics of the DNA double chain. In a pioneering paper appeared
in 1980, Englander, Kallenbach, Heeger, Krumhansl and Litwin
\cite{Eng} suggested that nonlinear excitations in the DNA double
chain could be instrumental in this process and allow the motion
of the transcription bubble to occur at near-zero energy cost.

In particular, as the fundamental motion undergone by DNA
nucleotides in this process is a roto/torsional one, they
suggested to model the DNA molecule as a double chain of coupled
pendulums; the relevant nonlinear excitations would then be
(topological) solitons pretty much like those well known in the
sine-Gordon equation. We refer to their work for detail.

Their suggestion was taken up by a number of author, producing
several variants of their fundamental model\footnote{In a related
but different direction, other authors also studied models for
radial stretching motions of the DNA molecule, which are related
in particular to DNA denaturation. The main line of research in
this direction followed the formulation of the Peyrard-Bishop
model \cite{PB} and extensions thereof \cite{BCP,BCPR,PeyNLN}. In
this note we will focus on roto/torsional dynamics.} (these are
reviewed in \cite{YakuBook}, to which the reader is referred for
detail and discussion). In particular, a simple model was put
forward by L.V. Yakushevich \cite{YakPLA}, see also
\cite{YakPhD,YakuBook,YakPRE} for further results and extensions;
this is recalled and discussed in sect.1 below.

The \Y (Y) model has its appeal, and at the same time weakness, in
its simplicity; it is remarkable that despite such simplicity it
predicts correct orders of magnitude for a number of physical
features, such as timescales of small oscillations and -- most
relevant in the wake of the approach of Englander {\it et al.} --
size of the nonlinear excitations: in fact, this corresponds to
the experimental size of the transcription bubble.

At the same time, the \Y model is a first-grade (i.e. one degree
of freedom per nucleotide \cite{YakPhD}), idealized (i.e. DNA is
considered as a homogeneous chain \cite{YakPhD}), and
nearly-linear model: interaction between successive nucleotides is
via a harmonic potential, and the intrapair interaction (i.e.
interaction between bases in a Watson-Crick pair, and though these
between the corresponding nucleotides) is also modelled by a
potential which is harmonic in the distance, and becomes
anharmonic when described in terms of the relevant degrees of
freedom, i.e. rotation angles.

While the first two features are nearly unavoidable (but see
\cite{SacSgu}) in models to be analyzed in analytical -- and not
just numerical -- terms, near-linearity raises more perplexity
even from the point of view of a mathematical approach, in
particular since the motions we are mainly interested in are fully
nonlinear.

In this note we consider a modified Y model, in which the
intrapair potential is no longer near-linear. Our main motivation
in this study was to understand how much of the prediction of Y
model survives a fully nonlinear modelling of the intrapair
potential. Note in this respect that the Peyrard-Bishop (PB) model
\cite{PB} considers a fully nonlinear Morse potential for this
interaction. In this case we should also consider the
directionality of H bonding, which plays no role in the
``straight'' motions considered in the PB model and is not taken
into account by the Morse-PB potential.

More radical modifications of the standard \Y model, modifying the
other features mentioned above, could also be considered; these
will be studied elsewhere \cite{CDG}.

As the interaction between bases in a pair is mediated by H bonds,
we argue that a more realistic approximation is that of a
potential describing H-bond interactions. This could be a Morse
potential as in the PB model, better if modified by taking into
account directionality effects; or even -- as we will do in the
first instance -- given that H bonds results from essentially
dipolar forces, a dipole-dipole interaction. This is not entirely
correct, since as bases rotate the charge distribution is also
modified, but takes into account several features of H bonding, in
particular its directionality, and of the motion geometry, which
are not considered in the standard Y model. In a way, the
dipole-dipole potential tests an approximation which is orthogonal
to the one represented by the \Y potential.

It should be mentioned that this work continues investigation
described in \cite{GaeYBC}, where we considered how the nonlinear
dynamics of the Y model is affected by dropping the contact
approximation $\ell_0 = 0$; as in the present case, we have shown
there that albeit the linear dynamics of the Y model is strongly
affected by the approximations this model assumes, the physical
features of fully nonlinear excitations are remarkably robust i.e.
survive when we drop those approximations (that of $\ell_0 = 0$ in
\cite{GaeYBC}, that of a harmonic intrapair potential here). The
same does not hold when one touches other features of the model,
as will be shown in forthcoming work \cite{CDG,SacSgu}.

In the following, after describing the standard Y model, including
also ``helicoidal'' terms \cite{Dauhel,Gaehel} (sect.1), we
discuss in some detail the case of general Y-like models and in
particular the reduced equations describing soliton solutions
(sect.2). These equations depend on the intrapair potential $V$
modelling H-bond interactions; we also focus on the special case
of soliton solutions of topological indices (1,0) and (0,1), see
below for definitions. In sections 3--5 we consider different
cases for the intrapair potential: the dipole-dipole potential in
sect.3, the Morse-PB potential in sect.4, and a Morse potential
modified by introducing a directional term in sect.5. In all
cases, we use the detailed discussion of sect.2 and tackle
directly the effective equations obtained there; these provide a
description of the special soliton equations which can be compared
-- as we do -- with those provided by the standard Y model. All
this discussion is conducted on continuous models, but the DNA
chain is intrinsically discrete; in sect.6 we evaluate numerically
a relevant parameter providing some measure of the relevance of
discreteness effects, i.e. the height of the Peierls-Nabarro
barrier. The final sect.7 is devoted to a discussion of the
results obtained and to drawing some conclusions.

\subsection*{Acknowledgements}

This work received support by the Italian MIUR (Ministero
dell'Istruzione, Universit\`a e Ricerca) under the program
COFIN2004, as part of the PRIN project {\it ``Mathematical Models
for DNA Dynamics ($M^2 \times D^2$)''}. I would like to thank M.
Cadoni, R. De Leo and G. Saccomandi for useful discussions.

\section{Yakushevich model}

The \Y model (Y model) is an ``ideal'' DNA model \cite{YakuBook}.
That is, the DNA double chain is considered to be of infinite
length, and all bases are considered as being equal.

In the \Y model, each nucleotide is considered as a single unit,
and its state described in terms of an angular variable. That is,
each nucleotide is seen as a rigid disk of radius $r$ which can
rotate around an axis and has a moment of inertia $I$ around this
axis; the rotation of the nucleotide at site $i \in {\bf Z}$ on
the chain $a$ ($a=1,2$) is described by an angle $\vth^{(a)}_i$.
When we are referring to a specific base pair, we write simply
$\vth_1 , \vth_2$ for $\vth^{(1)}_i,\vth^{(2)}_i$. The axes of
rotation of the two disks lie a distance $A \ge r$ away.

\begin{figure}
  \includegraphics[width=200pt]{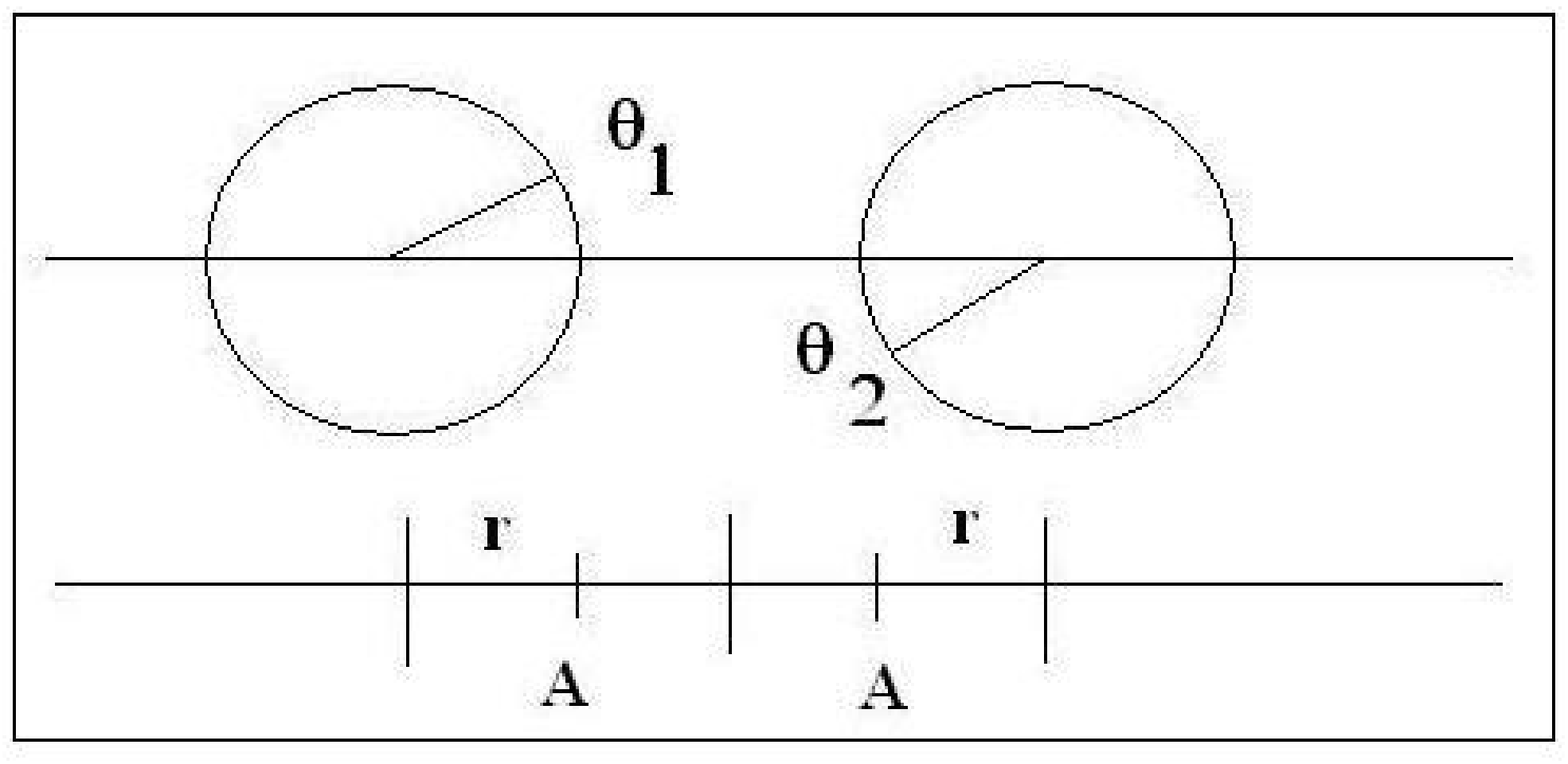}\ \hfill
    \includegraphics[width=80pt]{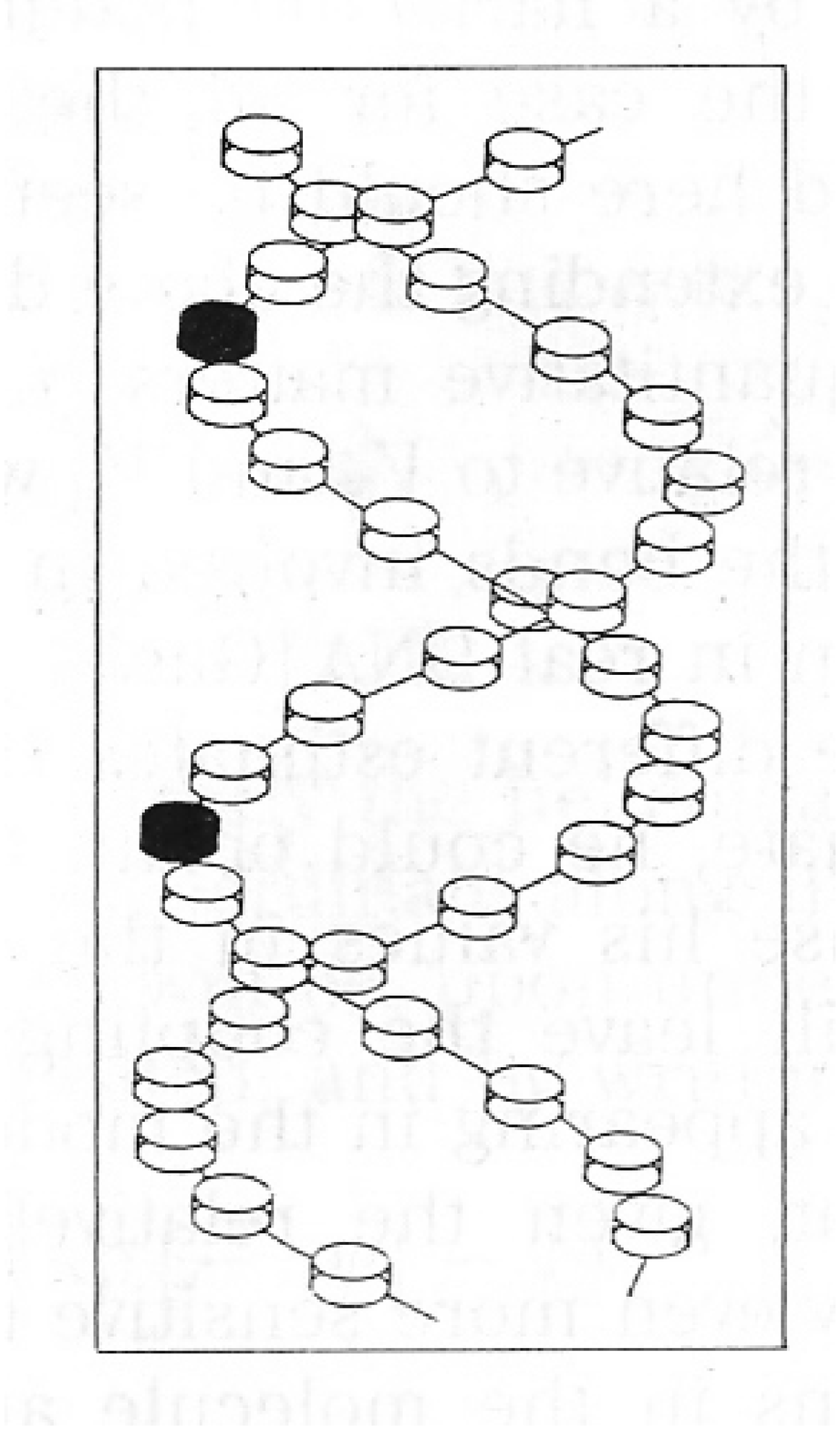}\\
  \caption{Left: A base pair in the Yakushevich model. Nucleotides are represented
  as disks of radius $r$; here $A = r + \ell_0 / 2$ is the distance between the centers of
  disks and the axis of the double helix.
  Right: Bases which are half-pitch of the helix away and on opposite chains
  are actually near enough in space, and Bernal-Fowler filaments \cite{Dav} joining them
  can form; they mediate the ``helicoidal interactions'' described by (1.3).}
\end{figure}

The model is better described by a Lagrangian $L = T - U$. In
this, the kinetic energy is obviously given by
$$  T \ = \ \sum_a \, \sum_i {I \over 2} \, ({\dot \vth}^{(a)}_i )^2 \ . \eqno(1.1) $$

The potential energy $U$ is the sum of two terms, or three if we
consider the so-called ``helicoidal'' version of the model
\cite{Dauhel,Gaehel,YakuBook}. These are the stacking ($U_s$)
term, the pairing one $(U_p$), and maybe the helicoidal one $U_h$;
the latter is relevant only in small amplitude dynamics and in the
dispersion relations. Thus $$ U \ = \  U_s \, + \, U_h \, + \, U_p
\ . $$

The first two terms correspond to harmonic potentials\footnote{We
stress that the choice of harmonic potentials for these term, and
in particular for the longitudinal interactions, is common to
nearly all DNA models \cite{PeyNLN,YakuBook}.} and depend on
dimensional coupling constants $K_s$ and $K_h$. The first,
describing backbone torsion and stacking interactions between
bases at nearby sites on the same chain, is given by
$$ U_s \ = \ {K_s \over 2} \ \sum_a \, \sum_i \, \( \vth^{(a)}_{i+1} -
\vth^{(a)}_{i} \)^2 \ . \eqno(1.2) $$

The helicoidal term describes interactions between bases which are
nearby in space due to helical geometry, and mediated in real DNA
by Bernal-Fowler filaments (chains of hydrogen bonded water
molecules \cite{Dav}), which join a base and the one on the other
chain half-pitch of the helix away, see fig.1b; in B-DNA, this
means a distance of $h=5$ sites along the DNA double chain. These
are very weak interactions, and are also modelled by a harmonic
potential:
$$ U_h \ = \ {K_s \over 2} \ \sum_a \, \sum_i \, \( \vth^{(\^a)}_{i+h} -
\vth^{(a)}_{i} \)^2 \ . \eqno(1.3) $$ This term has no effect on
fully nonlinear dynamics \cite{GRPD}, but introduces qualitative
changes in small amplitude dynamics and dispersion relations; we
will thus drop it when considering fully nonlinear excitations.

As for the $U_p$ term, it describes the nonlinear interactions
between bases in a pair, i.e. $$ U_p \ = \ \sum_i \ V(\vth^{(1)}_i
, \vth^{(2)}_i ) \eqno(1.4) $$ with $V$ the intrapair potential.

We will call the models of the class described so far, with
general $V$, \Y-like models; we denote the original \Y model (with
$V$ given by (1.5) below) as the ``standard'' Y-model.

In the standard Y model it is assumed that $$ V \ = \ (1/2) \, K_p
\ (\ell - \ell_0)^2 \ , \eqno(1.5) $$ where $\ell$ is the distance
between the atoms bridged by the H bonds, and $\ell_0$ is this
distance in the equilibrium configuration described by
$\vth^{(a)}_i = 0$. Note $r + \ell_0 /2 = A$.

This harmonic potential becomes anharmonic in terms of the angles
$\vth^{(a)}_i$. Indeed, with simple algebra (and writing, for ease
of notation, $\vth_a$ for $\vth^{(a)}_i$) we have\footnote{It
should be stressed that -- as remarked by Gonzalez and
Martin-Landrove \cite{GML} -- if we expand this in a power series
for $\vth_1 , \vth_2$ around $\vth_1 = \vth_2 = 0$, and assuming
$\ell_0 \not= 0$, the quadratic term vanishes and $ (\ell -
\ell_0)^2 \approx [r^2 / ( 4 \ell_0^2)] [ r (\vth_1 + \vth_2 )^2 +
\ell_0 (\vth_1^2 + \vth_2^2) ]^2$.} $ \ell^2 = (\ell_0^2 + 4
\ell_0 r + 6 r^2 ) - 2 r (\ell_0 + 2 r) [ \cos \vth_1 + \cos
\vth_2 ]  + 2 r^2 \cos (\vth_1 - \vth_2 )$.

In the usual treatments of the Y model, one considers the
approximation $\ell_0 = 0$, also called ``contact approximation''
(a discussion of the usual Y model beyond the contact
approximation is given in \cite{GaeYBC}). With this, $r \simeq A$
and
$$ (\ell - \ell_0)^2 \ = \ \ell^2 \ = \ 2 r^2 [ 3 - 2 (\cos(\vth_1) + \cos
(\vth_2) ) + \cos (\vth_1 - \vth_2 ) ] \ ; $$ expanding around the
equilibrium we get $ \ell^2 \approx r^2 ( \vth_1 + \vth_2 )^2$.
Thus,
$$ V \ = \ K_p r^2 \ \[ \cos (\vth^{(1)}_i -
\vth^{(2)}_i ) - 2 \, \( \cos \vth^{(1)}_i + \cos \vth^{(2)}_i \)
+ 3 \] \ . \eqno(1.6) $$ The intrapair potential term $U_p$ reads
from this and (1.4). This completes the description of the \Y
model.\footnote{In the following we will consider a different
expression for $V$, which we believe is physically better
justified than (1.5), which will also dispense us with the
prescription $\ell_0 = 0$. We will first describe other features
of the standard \Y model}.

In studying the Y model it is convenient to pass to variables
$$ \psi_n = \( \vth^{(1)}_n + \vth^{(2)}_n \) / 2 \ , \
\chi_n = \( \vth^{(1)}_n - \vth^{(2)}_n \) /  2 \eqno(1.5) $$
(thus $\vth^{(1)}_n = \psi_n + \chi_n$ and $\vth^{(2)}_n = \psi_n
- \chi_n$), and introduce
$$ a = K_s/I \ , \ b = K_h/I \ , \ c = 2 r^2 (K_p/I) \ . \eqno(1.6) $$

The Euler-Lagrange equations stemming from the \Y Lagrangian are
$$ \begin{array}{rl}
{\ddot \psi}_n \ =& \ a \, (\psi_{n+1} - 2 \psi_n + \psi_{n-1} ) \
+ \ b (\psi_{n+h} - 2 \psi_n + \psi_{n-h} ) \ + \\ & \ - \ c \, \sin \psi \, \cos \chi \ , \\
{\ddot \chi}_n \ =& \ a \, (\chi_{n+1} - 2 \chi_n + \chi_{n-1} ) \
- \ b (\chi_{n+h} + 2 \chi_n + \chi_{n-h} ) \ + \\ & \ - \ c \,
\sin \chi \, \( \cos \psi -  \cos \chi \) \ .
\end{array} \eqno(1.7) $$

Linearizing these about the equilibrium $\psi_n = \chi_n = 0$ and
writing $\psi_n = f_{q \om} \exp[i (q n + \om t)]$, $\chi_n = g_{q
\om} \exp[i (q n + \om t)]$, $q = k \de$, yields the dispersion
relations
$$ \begin{array}{l}
\om^2_\psi \ = \ 2 a \, (1 - \cos (k \de) ) \
+ \ 2 b (1 - \cos (k \de h) ) \ + \ c \ ; \\
\om^2_\chi \ = \ 2 a \, (1 - \cos (k \de ) ) \ + \ 2 b (1 + \cos
(k \de h) ) \ . \end{array} \eqno(1.8) $$

As for the dimensional parameters appearing in the model, we adopt
those suggested in \cite{GaeJBP}, i.e.
$$ \begin{array}{c}
r = A = 5.5 \, {\rm \AA} \ , \ \ \de = 3.4 \, {\rm \AA} \ ; \ \ I
= 3 \cdot 10^{-37} {\rm cm}^2\cdot{\rm g} \ ; \\
K_s = 0.13 {\rm eV/rad}^2 \ , \ K_p r^2 = 0.025 {\rm eV/rad}^2 \ ,
\ K_h = 0.009 {\rm eV/rad}^2 \ . \end{array} \eqno(1.9) $$

The choice for $K_p$ corresponds to an energy of $0.1 {\rm eV}$
for the breaking of the intrapair coupling (this is $\Delta E =
V(\pi,\pi) - V(0,0)$); this corresponds to 2.5 times the energy
needed for breaking a H bond, i.e. is intermediate between the A-T
and the G-C cases.

The full equations admit travelling wave solutions; these are
better discussed in the continuum approximation in which the
arrays $\{\vth^{(1)}_i,\vth^{(2)}_i \}$, and hence $\{\psi_i
(t)\}$ and $\{\chi_i (t)\}$, are promoted to fields $\vth_a
(x,t)$, $\phi (x,t)$, $\xi (x,t)$. The finite energy condition
require that each $\vth_a (x,t)$ goes to a multiple of $2 \pi$
(with vanishing $x$ derivative) for $|x|\to \infty$; hence finite
energy travelling wave solutions are indexed by two integer
rotation numbers $(m,n)$. It turns out that there is a maximum
speed $v_0 = \sqrt{a} \de$ for these.

The fundamental soliton solutions can be easily determined
analytically. They are given, respectively, by
$$ \begin{array}{ll}
\vphi (z) \ = \ 4 \, {\rm arctan} [ e^{\kappa z} ] \ ,& \ \xi = 0
\
; \\
\xi (z) \ = \ 2 \, {\rm arccos} \({ - \kappa z / \sqrt{1 +
\kappa^2 z^2} } \) \ ,& \vphi = 0 \ . \end{array} \eqno(1.10) $$
Here we have written $ \kappa := \sqrt{c / (a \de^2 - I v^2)}$;
for $v=0$ and with the above choices for the parameters, this
yields
$$ \kappa \ = \ \de^{-1} \, \sqrt{2 K_p r^2 / K_s} \ \approx \ 0.62
\, \de^{-1} \ \approx \ 0.18 \, {\rm \AA}^{-1} \ . \eqno(1.11) $$
We refer e.g. to \cite{GRPD,YakuBook} for further detail.

\section{General Y-like models}

As mentioned above, a general Y-like model is obtained when we
substitute (1.5) with a general intrapair potential
$V(\vth_1,\vth_2)$ having a minimum in $\vth_1=\vth_2=0$ (and only
there, up to $2 \pi$ periodicity), and symmetric under the
exchange of the two chains, $V(\vth_1,\vth_2) = V(\vth_2,\vth_1)$.

The analysis conducted for the standard \Y model can be to a
certain extent generalized to this wider class of models. From now
on, we will use the term ``\Y model'' in the generalized sense.

\subsection{Equations of motion}

The equations of motion -- i.e. the Euler-Lagrange equations
arising from the modified Lagrangian -- are
$$ \begin{array}{rl}
I {\ddot \vth}^{(a)}_n =& K_s \( \vth^{(a)}_{n+1} - 2 \vth^{(a)}_n
+ \vth^{(a)}_{n-1} \) + K_h \( \vth^{(\^a)}_{n+h} - 2 \vth^{(a)}_n
+ \vth^{(\^a)}_{n-h} \) \ + \\
& \ - \pa V(\vth^{(1)}_n,\vth^{(2)}_n) / \pa \vth^{(a)}_n \ .
\end{array} \eqno(2.1) $$
We now pass to the $\psi , \chi$ variables, and write $V_a$ for
$V$ expressed in terms of these to avoid any misunderstanding,
i.e. $V_a (\psi,\chi) = V (\psi + \chi, \psi-\chi)$. We get (note
the sign differences in the $K_h$ term and the $1/2$ factor in the
$V_a$ terms)
$$ \begin{array}{rl}
I {\ddot \psi}_n \ =& \ K_s \, \( \psi_{n+1} - 2 \psi_n +
\psi_{n-1} \) \ + \ K_h \( \psi_{n+h} - 2 \psi_n + \psi_{n-h} \) \
+ \\
 & \ - \ (1/2) \(\pa V_a(\psi,\chi) / \pa \psi \) \ , \\
I {\ddot \chi}_n \ =& \ K_s \, \( \chi_{n+1} - 2 \chi_n +
\chi_{n-1} \) \ - \ K_h \( \chi_{n+h} + 2 \chi_n + \chi_{n-h} \) \
+ \\
& \ - \ (1/2) \(\pa V_a(\psi,\chi) / \pa \chi \) \ . \end{array}
\eqno(2.2) $$

\subsection{Small oscillations}

In order to analyze small amplitude dynamics around the
$\vth^{(a)}_i = 0$ equilibrium, we consider the quadratic
approximation $V_0$ for $V$. Due to the symmetry property required
of $V$, see above, this will be decomposable as a symmetric and an
antisymmetric part for $\vth_1 \rightleftharpoons \vth_2$; hence
we can write $V_0 = c_+ (\vth_1 + \vth_2)^2 + c_- (\vth_1 -
\vth_2)^2$ or, in terms of the $\psi,\chi$ variables,
$$ V_{a,0} (\psi, \chi) \ = \ (1/2) [(8 c_+ ) \, \psi^2 \ + \
(8 c_-) \, \chi^2 \ . \eqno(2.3) $$

The dispersion relations are obtained setting $\psi_n (t) = f_{q
\om} \exp [i(q n + \om t)]$, $\psi_n (t) = g_{q \om} \exp [i(q n +
\om t)]$, and working at the linear level. In this way, (2.2)
yields
$$ \begin{array}{l}
\om^2_\psi \ = \ 2 a [1 - \cos (q) ] + 2 b [1 - \cos (q h)] +
c_\psi   \\
\om^2_\chi \ = \ 2 a [1 - \cos (q) ] + 2 b [1 + \cos (q h)] +
c_\chi \end{array} \eqno(2.4) $$ for the $\psi$ and the $\chi$
equations respectively; in these we have used $a$ and $b$ as
defined in (1.6), and introduced the constants
$$ c_\psi \ := \ 8 c_+ / I \ , \ \
c_\chi \ := \ 8 c_- / I \ . \eqno(2.5) $$

Note that the $\psi$ branch is acoustical (i.e. $\lim_{q \to 0}
\om (q) = 0 $) or optical (i.e. $\lim_{q \to 0} \om (q) \not= 0 $)
according to the vanishing or otherwise of the constant $c_\psi$,
while for the $\chi$ branch this depends on the constant $4 b +
c_\chi$.

\subsection{Continuum approximation}

We will now consider the full nonlinear dynamics (2.2). In the
study of the standard \Y model, the ``helicoidal'' term turns out
to be irrelevant in the fully nonlinear regime; on the other hand,
they make difficult a discussion of soliton solutions. We will
therefore drop these (i.e. set $K_h = 0$) from now on.

Having set $K_h=0$, eqs. (2.2) reduce to
$$ \begin{array}{rl}
I {\ddot \psi}_n \ =& \ K_s \, \( \psi_{n+1} - 2 \psi_n +
\psi_{n-1} \) \ - \ (1/2) \(\pa V_a(\psi,\chi) / \pa \psi \) \ , \\
I {\ddot \chi}_n \ =& \ K_s \, \( \chi_{n+1} - 2 \chi_n +
\chi_{n-1} \) \ - \ (1/2) \(\pa V_a(\psi,\chi) / \pa \chi \) \ .
\end{array} \eqno(2.6) $$

In discussing these, it is convenient to promote the arrays $\{
\psi_n (t) \}$ and $\{ \chi_n (t) \}$ ($n \in {\bf Z}$) to fields
$\Phi (x,t)$ and $\Xi (x,t)$, the correspondence being given by
$$ \psi_n (t) \ \simeq \ \Phi (n \de , t ) \ , \ \ \chi_n (t) \
\simeq \ \Xi (n \de , t) \ . \eqno(2.7) $$ Here $\de$ represents
the spacing between successive sites; in B-DNA, $\de = 3.4$ \AA.

With this, and $\^V = (1/2) V_a$, (2.6) reads
$$ \begin{array}{rl}
I \Phi_{tt} (x,t) =& K_s \( \Phi (x+ \de,t) - 2 \Phi (x,t)
+ \Phi (x - \de , t)  \) - \(\pa \^V (\Phi,\Xi) / \pa \Phi \) (x,t) \ , \\
I \Xi_{tt} (x,t) =& K_s \( \Xi (x+ \de,t) - 2 \Xi (x,t) + \Xi (x -
\de , t)  \) - \(\pa \^V (\Phi,\Xi) / \pa \Xi \) (x,t) \ .
\end{array} \eqno(2.8) $$

If now we assume that $\Phi$ and $\Xi$ vary slowly in space
compared with the length scale set by lattice spacing, we can
write
$$ \begin{array}{l}
\Phi (x \pm \de , t) = \Phi (x,t) \pm \de \Phi_x (x,t) + (\de^2 /
2) \Phi_{xx} (x,t) \ , \\
\Xi (x \pm \de , t) = \Xi (x,t) \pm \de \Xi_x (x,t) + (\de^2 / 2)
\Xi_{xx} (x,t) \ . \end{array} \eqno(2.9) $$

Inserting these into (2.8), we finally obtain the field equation
for the improved \Y model in the continuum approximation:
$$ \begin{array}{rl}
I \Phi_{tt} (x,t) \ =& \ K_s \, \de^2 \, \Phi_{xx} (x,t)
\ - \ \(\pa \^V (\Phi,\Xi) / \pa \Phi \) (x,t) \ , \\
I \Xi_{tt} (x,t) \ =& \ K_s \, \de^2 \Xi_{xx} (x,t) \ - \ \(\pa
\^V (\Phi,\Xi) / \pa \Xi \) (x,t) \ .
\end{array} \eqno(2.10) $$
We can from now on omit to write the point $(x,t)$ at which
functions should be evaluated, as all of them refer to the same
point.

The (2.10) being PDEs, they should be supplemented with a side
condition determining the function space to which the acceptable
solutions belong. The physically natural condition is that of {\it
finite energy}; that is, we should require that the integral
$$ \int_{-\infty}^{+\infty} \[ {1 \over 2} \( I (\Phi_t^2 + \Xi_t^2 ) + K_s
(\Phi_x^2 + \Xi_x^2) \) \ + \ \^V (\Phi , \Xi ) \] \ \d x
\eqno(2.11) $$ is finite; note that if this condition is satisfied
at $t=0$, it will be so for any $t$.

It should also be mentioned that considering a continuum version
of the model leads to disappearance of the Peierls-Nabarro barrier
\cite{PeyNLN} for soliton motion.

\subsection{Travelling wave solutions}

Next we focus on travelling wave solution for (2.10). That is, we
restrict (2.10) to a space of functions
$$ \Phi (x,t) \ = \ \vphi (x - v t) \ , \ \Xi (x,t) \ = \ \eta (x -
v t) \eqno(2.12) $$ (we should further restrict this in order to
take into account the finite energy condition). We will also write
simply $z := x - v t$, and introduce the parameter
$$ \mu \ := \ I \, v^2 \ - \ K_s \, \de^2 \ . \eqno(2.13) $$
With this, (2.10) reduce to two ODEs, i.e.
$$ \begin{array}{rl}
\vphi'' \ =& \ - \, \(\pa W (\vphi,\eta) / \pa \vphi \) \ , \\
\eta'' \ =& \ - \, \(\pa W (\vphi,\eta) / \pa \eta \) \ ,
\end{array} \eqno(2.14) $$
where we have of course defined
$$ W (\vphi , \eta ) \ := \ \mu^{-1} \ \^V (\vphi , \eta ) \ . $$
Tracing back the definition of $W$ in terms of the original
intrapair potential $V$, we obtain
$$ W(\vphi , \eta ) \ = \ {1 \over 2 \mu} \
V (\vphi+\eta,\vphi-\eta) \ . \eqno(2.15) $$

That is, travelling wave solutions are described by the motion of
a point particle of unit mass in the potential $W$, with $z$
playing the role of time for this motion. Note that $\mu$ could be
negative, which will actually be the relevant case. The
conservation of energy reads
$$ {1 \over 2} \( (\vphi')^2 + (\eta')^2 \) \ + \ W(\vphi ,
\eta ) \ = \ E \ . \eqno(2.16) $$

\subsection{Soliton solutions}

Let us come to the finite energy condition (2.11). In terms of our
functions $\vphi (z)$, $\eta (z)$, we require that
$$ \lim_{z \to \pm \infty} \vphi' (z) \ = \ \lim_{z \to \pm \infty} \eta' (z) \ =
\ 0 \ ; \eqno(2.17') $$ moreover, the functions $\vphi$ and $\eta$
themselves should go to a point of minimum for the potential
$\^V$.

Note that if $\mu >0$, the minima of $\^V$ are the same as the
minima of $W$; but if $\mu < 0$, then minima of $\^V$ are the same
as the maxima of $W$. As one cannot have nontrivial motions which
reach asymptotically in time (that is, in $z$) a minimum of the
effective potential, in order to have travelling wave solutions
satisfying (6.6') and going to minima of $\^V$ for $z \to \pm
\infty$ we need $\mu < 0$; we assume this from now on, and write
$$ \mu \ = \ - \, 1/\tau^2 \ . \eqno(2.18) $$
In turn, $\mu < 0$ implies that there is a maximum speed for
travelling waves:
$$ |v| \ < \ (\sqrt{K_s / I }) \, \de \ := \ v_{max} \ . \eqno(2.19)
$$

The minima of $\^V$, i.e. the maxima of $W$, are for $\vth_i = 2
\pi \s_i$, i.e. $ \vphi = 2 n \pi$, $\eta = 2 m \pi$ with $m,n$
half-integers, $n = (s_1 + s_2)/2$, $m = (s_1 - s_2)/2$; thus the
finite energy condition requires
$$ \begin{array}{l}
\lim_{z \to \pm \infty} \vphi (z) \ = \ 2 n_\pm \pi \ , \ \lim_{z
\to \pm \infty} \eta (z) \ = \ 2 m_\pm \pi \ ; \\
n_\pm = {1 \over 2} (s_1^\pm + s_2^\pm) \ , \ m_\pm = {1 \over 2}
(s_1^\pm - s_2^\pm ) \ . \end{array} \eqno(2.17'')
$$ The integers
$ \s_i := s_i^+ - s_i^-$ are rotation numbers, counting the number
of complete turns made by the angles $\vth_i$ in between $z = -
\infty$ and $z = + \infty$; they thus represent a topological
index and separate functions satisfying the finite energy
condition into distinct topological sectors. The same applies to
$$ n := n_+ - n_- \ \ {\rm and} \ \ m := m_+ - m_- $$
(we can always take $n_- = m_- = 0$ by suitably choosing the
origin for the angles $\vphi$ and $\eta$). The solutions with
nonzero $(m,n)$ will correspond to topological solitons. As the
problem admits a variational formulation, we are sure there are
solutions in each topological sector \cite{DNF}.

In terms of the dynamical system describing the evolution in
``time'' $Z$ in the potential $W$, they represent heteroclinic
solutions connecting the point $(0,0)$ at $z = - \infty$ with the
point $(2 \pi n , 2 \pi m )$ at $z = + \infty$. It is thus no
surprise that the analytic determination of such solutions is in
general impossible.

\subsection{Special soliton solutions}

The soliton solutions with indices $(1,0)$ (i.e. $\s_1 = \s_2 =
1$) and $(0,1)$ (i.e. $\s_1 = 1 = - \s_2$) are special and can be
determined explicitly.

These are special in that they make vary only one of the two
fields. That is, the $(1,0)$ solution will have $\eta (z) \equiv
0$, and the $(0,1)$ solution will have $\vphi (z) \equiv 0$. Thus,
they correspond to one-dimensional motions in the effective
potential $W (\vphi,\eta)$, and as such they can be exactly
integrated, as we now discuss.

By adding a constant we can always set $V(0,0) = 0$ and hence
$W(0,0) = 0$, see (2.15), so we assume this to be the case.

The (1,0) soliton corresponds to $\eta=0$, so we define
$$ \^P(\vphi ) \ := \ W(\vphi,0) \ , \eqno(2.20) $$
and rewrite the conservation of energy (2.16) as
$$ {1 \over 2} \( {\d \vphi \over \d z} \)^2 \ + \ \^P (\vphi ) \ = \
E \ . \eqno(2.21) $$ We are interested in the motion with $E = E_0
= W(0,0)$; we assumed this to be $E_0=0$.

Eq.(2.21) yields then the separable equation
$$ {\d \vphi \over \d z} \ = \ \sqrt{ - 2 \, \^P(\vphi ) } \ ,
\eqno(2.22') $$ or equivalently
$$ {\d \vphi \over \sqrt{2 \^P (\vphi )}} \ = \ \d z \ . \eqno(2.22'')
$$
Introducing the reduced effective potential (the second equality
follows from (2.15))
$$ P(\vphi) \ := \ 2 \mu \ \^P(\vphi) \ = \ V (\vphi , \vphi )
\eqno(2.23) $$ and recalling (2.18), we get immediately the
equation for the (1,0) soliton:
$$ {\d \vphi \over \sqrt{P(\vphi)}} \ = \ \tau \, \d z \ . \eqno(2.24) $$

Denoting by $f$ the integral of the l.h.s.,
$$ f (\vphi) \ := \ \int {\d \vphi \over \sqrt{P(\vphi )}} \ ,
\eqno(2.25) $$ we get hence
$$ f (\vphi ) = \tau (z - z_0) \ , \ \ \vphi = f^{-1} [\tau (z - z_0)] \ .
\eqno(2.26) $$

The same approach allows to obtain the (0,1) soliton, for which
$\vphi = 0$. In this case we define
$$ \^Q(\eta ) \ := \ W(0,\eta) \ , \eqno(2.27) $$
and rewrite the conservation of energy (2.16) as
$$ {1 \over 2} \( {\d \eta \over \d z} \)^2 \ + \ \^Q (\eta ) \ = \
E \ . \eqno(2.28) $$ We again consider $E=E_0=0$; eq.(2.28) yields
then
$$ {\d \eta \over \sqrt{2 \^Q(\eta )}} \ = \ \d z \ , \eqno(2.29) $$
and introducing, see again (2.15) for the second equality,
$$ Q(\eta) \ := \ 2 \mu \ \^Q(\eta) \ = \ V (\eta , - \eta ) \ ,
\eqno(2.30) $$ we get immediately the equation for the (0,1)
soliton:
$$ {\d \vphi \over \sqrt{Q(\vphi)}} \ = \ \tau \, \d z \ . \eqno(2.31) $$

Denoting by $g$ the integral of the l.h.s.,
$$ g (\eta) \ := \ \int {\d \eta \over \sqrt{Q(\eta )}} \ ,
\eqno(2.32) $$ we get hence
$$ g (\eta ) = \tau (z - z_0) \ , \ \ \eta = g^{-1} [\tau (z - z_0)] \ .
\eqno(2.33) $$

It may be worth recalling, as a final remark in our general
discussion, that in view of (2.23) and (2.30), the equations
(2.22) and (2.31) obeyed by the (1,0) and the (0,1) solitons are
also written in terms of the original intrapair potential $V
(\vth_1 , \vth_2)$ as
$$ \d \vphi / \d \zeta \ = \ \sqrt{V(\vphi , \vphi)} \eqno(2.34) $$
for the (1,0) soliton, and
$$ \d \eta  / \d \zeta \ = \ \sqrt{V(\eta , - \eta)} \eqno(2.35) $$
for the (0,1) soliton. Here we introduced the rescaled variable $$
\zeta \ := \ \tau \ z \ = \ z / \sqrt{|\mu |} \ = \ z / (\de
\sqrt{K_s - I (v/\de)^2 }) \ ; \eqno(2.36) $$ the latter equality
follows from (2.13).


\section{Improving the intrapair potential I. \\
Dipole-dipole interaction}

The intrapair potential $V$ is due to the H bonds between
complementary bases. As H bonding is ultimately a dipolar
interaction, on a physical basis it appears preferable to consider
$V$ as a dipole-dipole potential rather than simply a harmonic
potential depending on variations of the euclidean distance of the
involved atoms.

Our first improved model will therefore be described as follows:
on the edge of the disks representing nucleotides in the \Y model
sits dipole of fixed strength and whose orientation follows the
bases orientation. For $\vth_1 = \vth_2 = 0$, the dipole have the
same orientation in space; as the angles change, they at the same
time acquire discordant orientation and vary their mutual
distance. The potential corresponding to given angles can be
easily computed by elementary trigonometry.

\subsection{Intrapair potential}

We consider an intrapair potential due to a dipole-dipole
interaction; this is obtained by expanding the full dipole-dipole
interaction potential -- obtained proceeding as described above --
in terms of the dipole momentum and keeping only second order
terms. Proceeding in this way we obtain, with $\ga$ a dimensional
parameter and $c_0$ an arbitrary constant
$$\begin{array}{rl}
V \ =& \ c_0 \ + \ \ga \ \[ 14 A r ( \cos \vth_1 + \cos \vth_2 ) +
2 A r
[ \cos (\vth_1 - 2 \vth_2 ) + \cos (\vth_2 - 2 \vth_1 ) ] + \right. \\
 & \left. - 4 (A^2 + 2 r^2 ) \cos (\vth_1 - \vth_2 ) - 12 A^2 \cos (\vth_1 +
\vth_2 ) + \right. \\
 & \left. - r^2 \cos (2 (\vth_1 - \vth_2 ) ) - 7 r^2 \]
\times
\\
& \times \[ 2 \sqrt{2} \( 2 A^2 + r^2 (1 + \cos (\vth_1 - \vth_2
)) - 2 A r ( \cos \vth_1 + \cos \vth_2 ) \)^{5/2} \]^{-1} \ .
\end{array} \eqno(3.1)
$$
We are of course assuming that $A \not= r$, i.e. we are {\it not }
adopting the \Y approximation $\ell_0 = 0$.

The above will be our choice for the intrapair potential, hence we
will have
$$ U_p \ = \ \sum_i \, V(\vth^{(1)}_i , \vth^{(2)}_i ) \ .
\eqno(3.2) $$ As for the other potential terms, we retain
Yakushevich's expressions.

In terms of the $\psi$ and $\chi$ variables defined above, $V$
reads
$$ \begin{array}{rl}
V_a \ =& c_0 \ - \ \ga \ \[7 r^2 + 12 A^2 \cos (2 \psi )
+ 4 (A^2 + 2 r^2 ) \cos (2 \chi ) + r^2 \cos (4 \chi ) + \right. \\
& \left. - 16 A r \cos \psi \cos \chi \( 1  + \cos^2 \chi \)  \] \times \\
& \times \[ 16 \, \( A^2 + r^2 \cos^2 \chi - 2 A r \cos \psi \cos
\chi  \)^{5/2} \]^{-1} \ .
\end{array} \eqno(3.3) $$

The potential is plotted in fig.2. Note that it has the properties
we should expect from a potential representing H bonding: it is
strongly directional, and the atoms are essentially free when
their position is not near enough to the equilibrium position.

The potential $V$ (or $V_a$) depends on three dimensional
parameters, i.e. the coupling constant $\gamma$ and the two
distances $A$ and $r$. It may be convenient -- in particular, in
order to compare the results for this model with those for the
standard \Y model -- to write it in terms of $\ga$, $A$ and the
adimensional constant $\la = r /A \le 1$. In this way, (3.1) reads
$$\begin{array}{rl}
V \ =& \ c_0 \ + \ \ga \ \, \[ 2 \la [ 7 (\cos \vth_1 + \cos
\vth_2) +  \cos (\vth_1 - 2 \vth_2 ) + \cos (\vth_2 - 2 \vth_1 ) ] + \right. \\
 & \left. - 4 (1 + 2 \la^2 ) \cos (\vth_1 - \vth_2 ) - 12 \cos (\vth_1 +
\vth_2 ) + \right. \\
 & \left. - \la^2 [7 + \cos (2 (\vth_1 - \vth_2 ) )] \]
\times
\\
& \times \[ 2 \sqrt{2} A^3 \( 2 + \la^2 (1 + \cos (\vth_1 - \vth_2
)) - 2 \la ( \cos \vth_1 + \cos \vth_2 ) \)^{5/2}
\]^{-1} \ . \end{array} \eqno(3.4) $$

We also note that in order to have $V(0,0) = 0$, the additive
constant $c_0$ should be chosen as
$$ c_0 \ = \ - \, \ga / [A (1 - \la)]^3  \ . \eqno(3.5) $$

\begin{figure}
  \includegraphics[width=150pt]{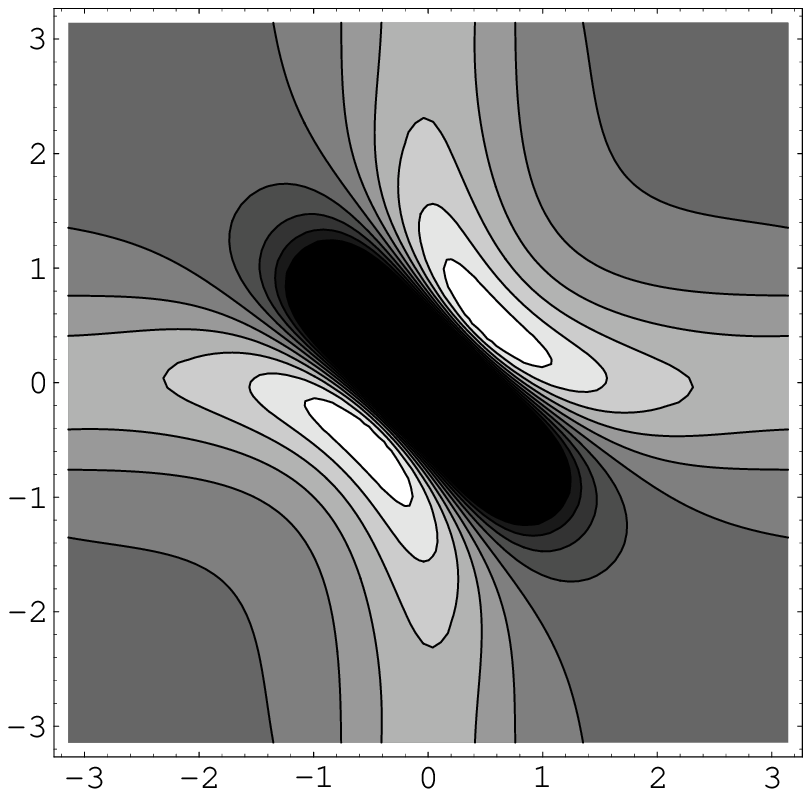} \ \
  \includegraphics[width=150pt]{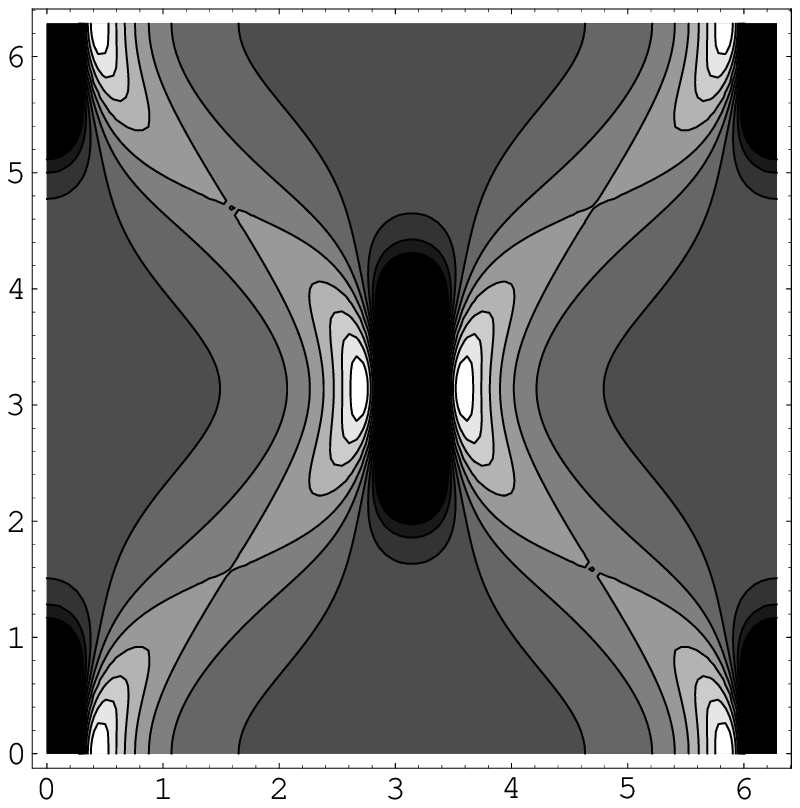}\\
  \caption{Left: The potential $V(\vth_1,\vth_2)$ as given by (3.1).
  Here we set $\ga = 1$
  and use the geometrical parameters $A = 5.5$ \AA, $r = 4.0$ \AA.
  Darker areas correspond to lower potential energy.
  Right: The intrapair potential $V_a(\psi,\chi)$ in terms of $\psi$
  and $\chi$ (parameters as above). Note that beside the $2 \pi$ periodicity
  in $\psi$ and in $\chi$, expressed by $V_a(\psi,\chi) = V_a(\psi + 2 \pi,\chi) =
  V_a(\psi,\chi + 2 \pi)$, there is also a periodicity for simultaneous
  $\pi$ translations in
  $\psi$ and $\chi$, i.e. $V_a(\psi,\chi) = V_a(\psi + \pi,\chi + \pi)$;
  this follows immediately from (1.5).}
\end{figure}

\subsection{Model parameters}

Our model depends on several parameters. Some of these -- i.e.
$A$, $K_s$ and $K_h$ are common to the standard \Y model, and we
will retain the values given in sect.1 for these.

Note that now $r$ is no longer coinciding with $A$; as the length
of H bonds is $\ell_0 \simeq 3 {\rm \AA}$ and $2 r = 2 A -
\ell_0$, we take $r = 4.0 {\rm \AA}$; that is, we choose
$$ r = 4.0 {\rm \AA} \ , \ \ell_0 = 3.0 {\rm \AA} \ , \ A = 5.5 {\rm
\AA} \ ; \ \ \ \la = (r / A) \ \simeq \ 0.7272... \ . \eqno(3.6)
$$

In order to select a value for $\ga$, we note that the equilibrium
corresponds to $\vth_1 = 0 = \vth_2$, while the ``maximally open''
state of the base pair is obtained for $\vth_1 = \pi = \vth_2$.
Thus $V(\pi,\pi) - V(0,0)$ should be of the order of the energy
$\Delta E$ needed to break the base pair; with the choice already
accepted for the standard \Y model, we should thus have
$V(\pi,\pi) - V(0,0) = \Delta E \simeq 0.1 {\rm eV}$.

On the other hand, (3.1) yields
$$ V(0,0) \ = \ c_0 \ - \ \ga / (A-r)^3 \ , \ \
V(\pi,\pi) \ = \ c_0 \ - \ \ga / (A+r)^3 \ ; \eqno(3.7) $$ hence
$$ V(\pi , \pi ) - V (0,0) = \gamma \( {1 \over (A - r)^3 }  - {1
\over (A+r)^3} \) = {\gamma \over A^3} \( {1 \over (1 - \la)^3 }
 - {1 \over (1+\la)^3} \)\ ; \eqno(3.8) $$ thus $\ga$ should be chosen
according to
$$ \ga \ = \ (\Delta E ) \ {(A^2 - r^2)^3 \over 2 r (3 A^2 + r^2 )}
\ = \ (\Delta E ) \ A^3 \ {(1 - \la^2)^3 \over 2 \la (3 + \la)} .
\eqno(3.9) $$ The relevant parameter for $V$ is $\ga /4 A^3$, and
by (3.9) we have
$$ {\ga \over 4 A^3} \ = \ {(1 - \la^2)^3 \over
8 \la (3+\la)} \ (\Delta E ) \ . \eqno(3.10) $$

With the values given above for $\Delta E$, $A$ and $r$, this
yields
$$ \ga \approx 0.34 \, {\rm \AA^3 \, eV} \ ; \ \ \
{\gamma \over 4 A^3} \ \approx \ 5.1 \cdot 10^{-4} \ {\rm eV} \ . \eqno(3.11) $$

\subsection{Special soliton solutions}

As discussed in sect.2, soliton solutions are obtained as motions
in the effective potential $W(\vphi,\eta)$, see (2.14); in the
present case $W$ is given -- as follows from (2.15) -- by
$$ \begin{array}{rl}
W (\vphi,\eta) \ =& - \, c_0 / (2 |\mu| ) \ + \\
 & + \  [\ga/(2 |\mu| A^3)] \, \[7 \la^2 +
12 \cos (2 \psi ) + 4 (1 + 2 \la^2 ) \cos (2 \chi )  + \right. \\
& \left. + \la^2 \cos (4 \chi )
- 16 \la \cos \psi \cos \chi \( 1  + \cos^2 \chi \)  \] \times \\
& \times \[ 16 \, \( 1 + \la^2 \cos^2 \chi - 2 \la \cos \psi \cos
\chi  \)^{5/2} \]^{-1} \ .
\end{array} \eqno(3.12) $$
It is convenient, for the computations to follow, to set $c_0$ as
given in (3.5).

\begin{figure}
  \includegraphics[width=150pt]{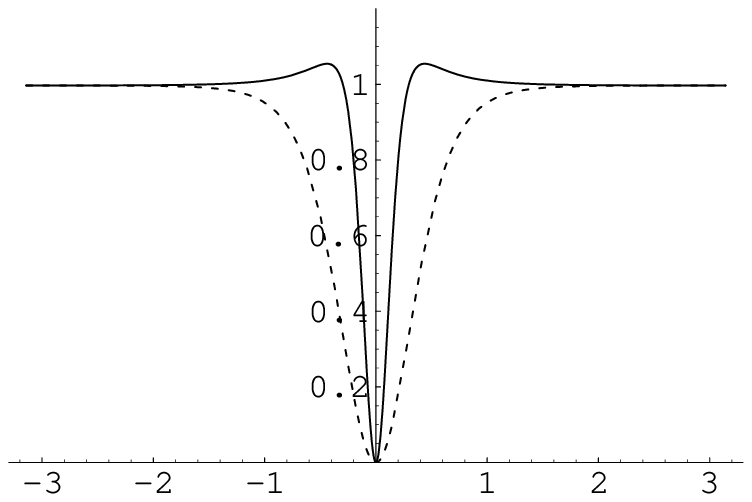}\ \
  \includegraphics[width=150pt]{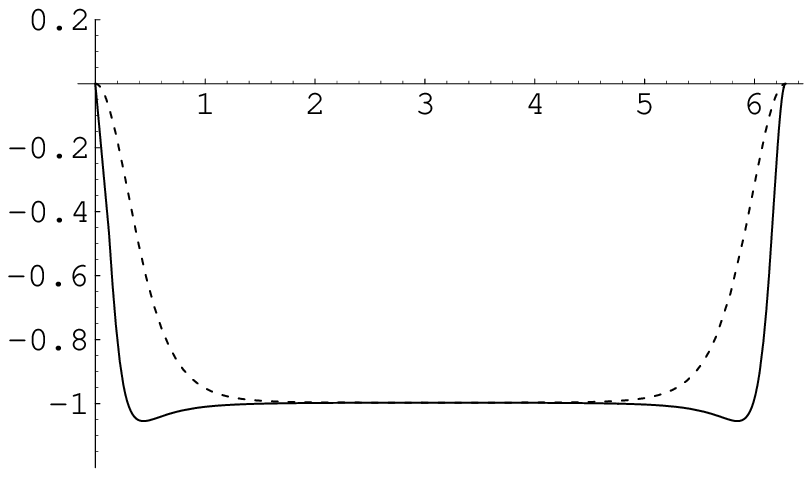}\\
  \caption{On the left: The potential $V_a (\psi,\chi)$ as a function of $\psi$ for
  $\chi=0$ (solid line) and as a function of $\chi$ for $\psi = 0$ (dashed line);
  we have used the values $A = 5.5$ \AA, $r = 4$ \AA and set $\mu = - 1$.
  On the right: The effective potentials $\^P (\vphi) = W (\vphi,0)$
  (solid) and $\^Q(\eta ) = W(0,\eta)$ (dashed) entering in the determination of
  the (1,0) and the (0,1) solitons. The additive constant $c_0$ has
  been chosen here so to have $V(0,0)=0$,
  and energy is measured in units such that $V(\pi,\pi) - V(0,0)=\Delta E = 1$.}
\end{figure}

\subsubsection{The (1,0) soliton}

The (1,0) soliton is obtained, as discussed in sect.2, by
considering $ P(\vphi) := V (\vphi,\vphi) $ and solving (2.24). In
the present case it results
$$ P (\vphi ) \ = \ {\ga \over 4 A^3 } \ \[ {4 \over (1 - \la )^3} \
- \ { 1 + 3 \cos (2 \vphi ) - 8 \la \cos (\vphi ) + 4 \la^2 \over
(1 - 2 \la \cos \vphi + \la^2 )^{5/2} } \] \ , \eqno(3.13) $$
where we have also used (3.5); we have then to solve

We were not able to perform the integration (2.24) analytically.
On the other hand the equivalent equation
$$ {\d \vphi \over \d z } \ = \
\tau \ \sqrt{P (\vphi)} \ , \eqno(3.14) $$ can be integrated
numerically. The result of this integration is shown in fig.4a,
where it is also compared with the standard \Y (1,0) soliton.

\subsubsection{The (0,1) soliton}

The (0,1) soliton is obtained, as discussed in sect.2, by
considering $ Q(\eta) := V (\eta,- \eta) $ and solving (2.31). In
the present case it results
$$ Q (\eta ) \ = \ {\ga \over 4 A^3 } \ \[ {4 \over (1 - \la )^3} \
- \ { 2 (1 + \cos^2 \eta) \over (1 - 2 \la \cos \eta + \la^2
)^{5/2} } \] \ , \eqno(3.15)
$$ where we have also used (3.5). Again we were not able to perform
the integration (2.31) analytically, but the equivalent equation
$$ {\d \eta \over \d z }  \ = \
\tau \ \sqrt{Q (\eta) } \eqno(3.16) $$ can be integrated
numerically. The result of this integration is shown in fig.4b,
where it is also compared with the standard \Y (0,1) soliton.

\begin{figure}
  \includegraphics[width=150pt]{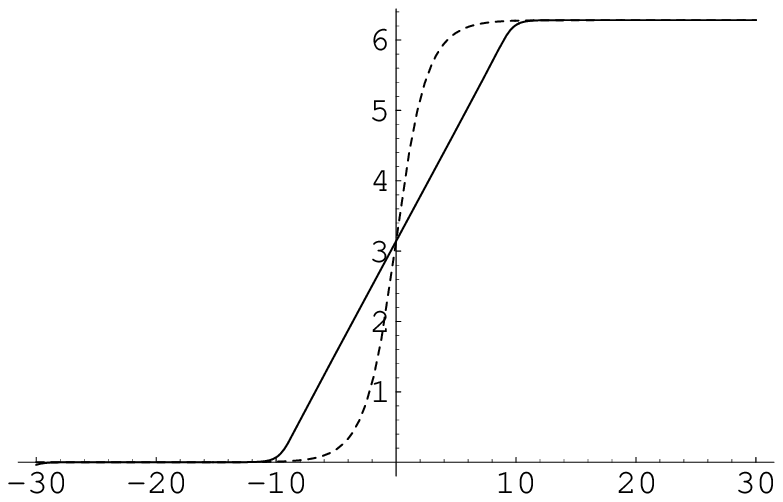}\ \
  \includegraphics[width=150pt]{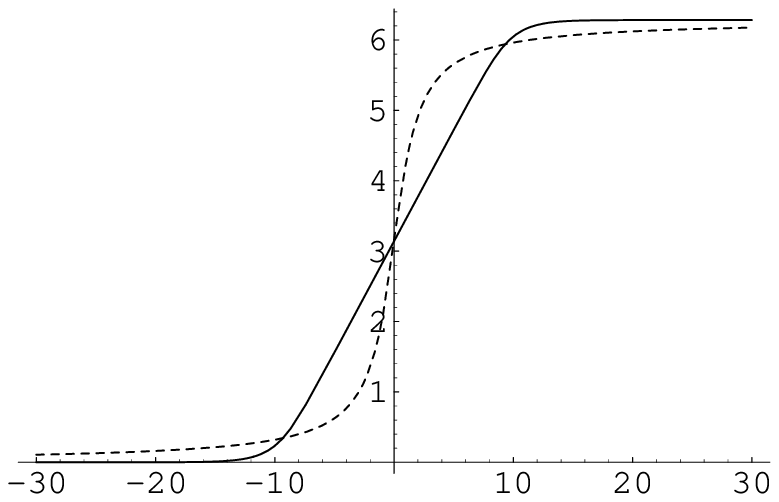}\\
  \caption{Comparison of the (1,0) (left) and the (0,1) (right) solitons for the
  standard \Y model (dotted) and our modified model (solid lines). Despite a
  certain difference in shape, the size of the (1,0) soliton -- corresponding to
  symmetric motions in the two helices -- is remarkably similar in the two models.}
\end{figure}

\section{Improving the intrapair potential II. \\
Morse potential}

The problem of modelling intrapair interactions mediated by H
bonds has been tackled in all attempts to provide dynamical models
of DNA; among these, the Peyrard-Bishop model (and later
improvements by Dauxois \cite{Dauhel} and by Barbi, Cocco and
Peyrard \cite{BCP,BCPR}) has a special relevance in view of its
success in describing DNA denaturation \cite{PeyNLN}. In this,
intrapair interactions are described by means of a Morse potential
$$ V (r) \ = \ \a \ \[ \(e^{- k (r - r_0)} - 1 \)^2 - 1 \]  \eqno(4.1) $$
where $r$ is the distance between (reference atoms belonging to)
bases in the given pair.

Note that $V$ has a minimum at $r=r_0$ with $V(r_0) = - \a$ and
$\lim_{r\to \infty} V(r) = 0$, so that $\a$ represents the binding
energy. We also have $V'' (r_0) = 2 \a k^2$.

\begin{figure}
  \includegraphics[width=150pt]{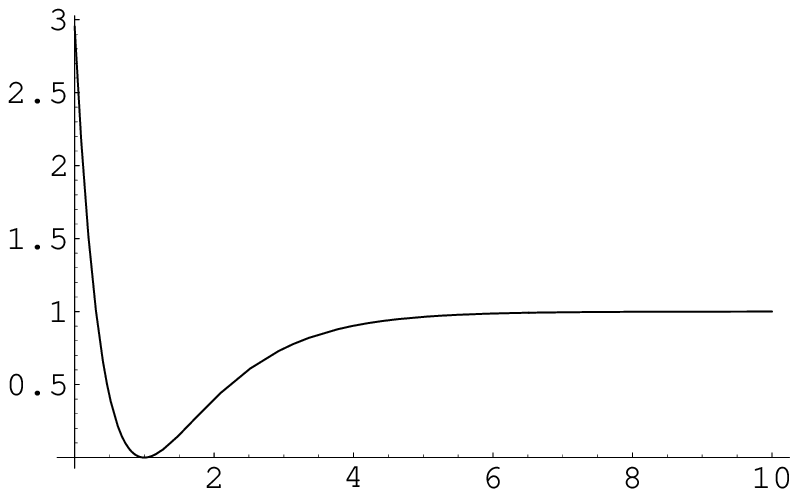} \hfill
  \includegraphics[width=150pt]{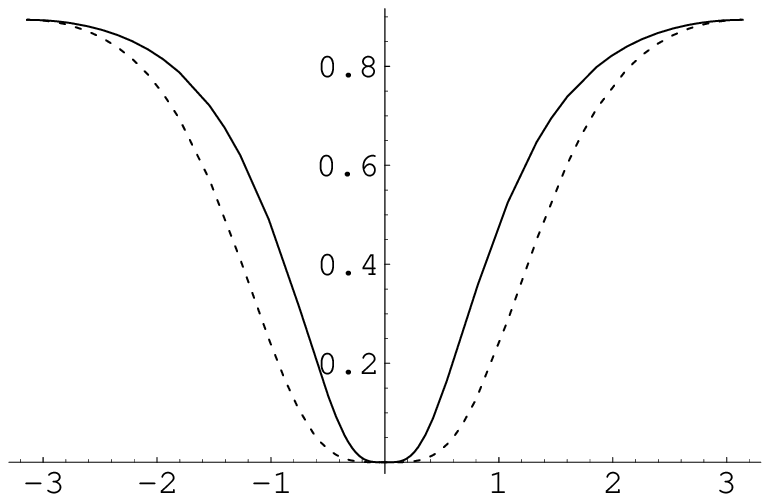}\\
  \caption{Left: The Morse-PB potential (4.1); here $D=1$, $k=1$ $x_0=1$,
  and we add a constant so that $V(x_0)=0$. Right: the resulting potential
  in terms of the rotation angle $\vth$, for $\vth_1 = \vth_2$ (solid) and for
  $\vth_1 = - \vth_2$ (dashed).}
\end{figure}

Thus, albeit the Peyrard-Bishop (PB) model aims at describing a
different process undergone by DNA, it is quite natural to
consider their modelization of the intrapair potential.

It should be noted that -- as pointed out in (4.1) -- the Morse-PB
potential depends only on the distance between H-bonded atoms, and
not on their alignement. This is no surprise, since in the PB
model and later improvements the bases lie on a straight line
passing through the axis of the double helix.\footnote{Moreover,
they are always assumed to move symmetrically. Asymmetrical
motions are also possible dynamically, but less favorable
energetically; moreover the dynamics is naturally split into
symmetrical and antisymmetrical motion, due to an exchange
symmetry. Thus one usually focuses on the more relevant sector of
symmetrical motions \cite{GRPD}.}

In our case we have to consider rotational motions, and thus the
latter conditions is not holding any more. This raises the
question of modelling angular dependence of H bonds in this
context; in this section we will consider the isotropic Morse-PB
potential (4.1), while in the next section we introduce an angular
dependence.

Also, for our purposes it is more convenient to set additive
constants in $V$ so that $V(0,0)=0$. Doing this, and denoting as
in previous section the distance between (relevant atoms in) the
two bases of the pair as $\ell$, and the distance equilibrium as
$\ell_0$, the Morse-PB potential reads
$$ V (\vth_1,\vth_2) \ = \
\a \ \(e^{- k (\ell(\vth_1,\vth_2) - \ell_0)} - 1 \)^2 \ .
\eqno(4.2) $$ Here we have pointed out explicitly the dependence
on the angles $\vth_i$. We also write, for ease of notation,
$$ \rho (\vth_1,\vth_2) \ := \ \ell (\vth_1,\vth_2) \, - \, \ell_0
\ , \eqno(4.3) $$ so that (4.2) reads
$$ V (\vth_1,\vth_2) \ = \
\a \ \(e^{- k \rho (\vth_1,\vth_2) } - 1 \)^2 \ . \eqno(4.4)
$$

\subsection{Soliton solutions}

With our choice of variables, the distance $\ell$ is written as
$$ \ell \ = \ 2 A \[ 1 - \la (\cos \vth_1 + \cos \vth_2 ) +
(\la^2 /2) \( 1 + \cos (\vth_1 - \vth_2) \) \]^{1/2} \eqno(4.5) $$
and of course $\ell_0 = 2 A (1 - \la ) $.

By standard computation,
$$ \begin{array}{l}
\rho (\vphi,\vphi) \ = \ 2 \, A \ \( \sqrt{1 - 2 \la \cos \vphi +
\la^2 } \, - \, (1 - \la) \) \ , \\
\rho (\eta,- \eta) \ = \ 2 \, A \ \( \sqrt{1 - 2 \la \cos \eta +
\la^2 \cos^2 \eta} \, - \, (1 - \la) \) \ . \end{array} \eqno(4.6)
$$

This and (4.4) immediately provide the symmetric and antisymmetric
reductions, $P(\vphi)$ and $Q(\eta)$, of the intrapair potential
which determine the (1,0) and (0,1) solitons via (2.34) and
(2.35); see (2.10) below.

\subsection{Model parameters}

The intrapair Morse-PB potential $V_p$ and hence the effective
potential $W$ depend on the two parameters $\a$ and $k$; this is
at variance with the Yakushevich potential and the dipole-dipole
potential considered in sect.3: they both depend on a single
parameter.

We can therefore impose two physical conditions in order to fix
these parameters; we will retain the values adopted above for the
other parameters.

As pointed out above, $\a$ represents the dissociation energy for
the Morse interaction, supposed to model the H-bond; hence we
should have
$$ \a \ = \ \Delta E \ \simeq \ 0.1 \, {\rm eV} \ . \eqno(4.7) $$

We also pointed out that for $\ell \simeq \ell_0$, $ V \simeq \a
k^2 (\ell - \ell_0 )^2$; comparing this with (1.5), we require
that
$$ 2 \, \a \, k^2 \ = \ K_p \ . \eqno(4.8) $$
Here $K_p$ should be chosen as in (1.10), taking into account our
choice for $r$. This yields
$$ k \ = \ \sqrt{K_p / (2 \a) } \ = \ r^{-1} \, \sqrt{K_p r^2 / (2 \Delta E)}
\ \simeq \ 0.09 \, {\rm \AA}^{-1} \ . \eqno(4.9) $$

\subsection{Special soliton solutions}

The relevant reduced potentials are now
$$ \begin{array}{l}
P (\vphi ) \ = \ V(\vphi,\vphi) \ = \ \a \, \( \exp \[ - 2 A k
\la (\sqrt{1 - 2 \la \cos \vphi } - (1 - \la) ) \] - 1 \)^2 \ ; \\
Q (\eta ) \ = \ V (\eta, - \eta) \ = \ \a \, \( \exp \[ - 2 A k
\la (1 - \cos \eta ) \] \)^2 \ . \end{array} \eqno(4.10) $$

The equations (2.34) and (2.35) governing the special (1,0) and
(0,1) soliton solutions are again impossible to solve
analytically, but can be integrated numerically. The results of
these integration are shown in fig.6, where we also plot the
standard \Y solitons of the same topological number for
comparison. Quite remarkably, again the overall size is about the
same as with the standard \Y model.

\begin{figure}
  \includegraphics[width=150pt]{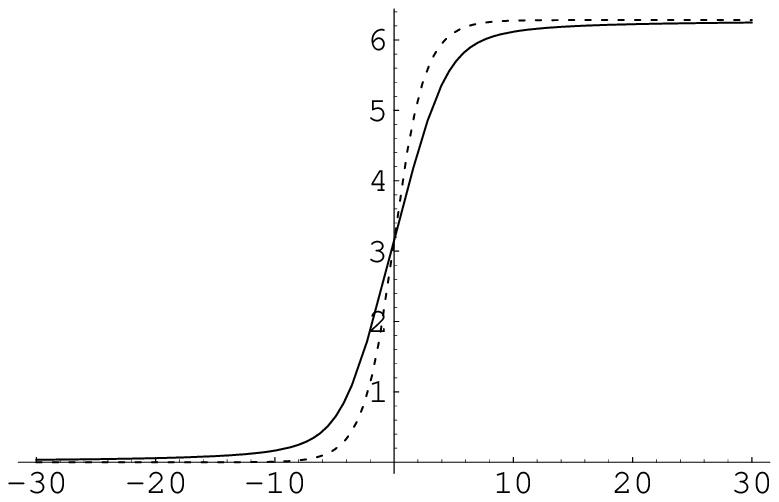}\
    \includegraphics[width=150pt]{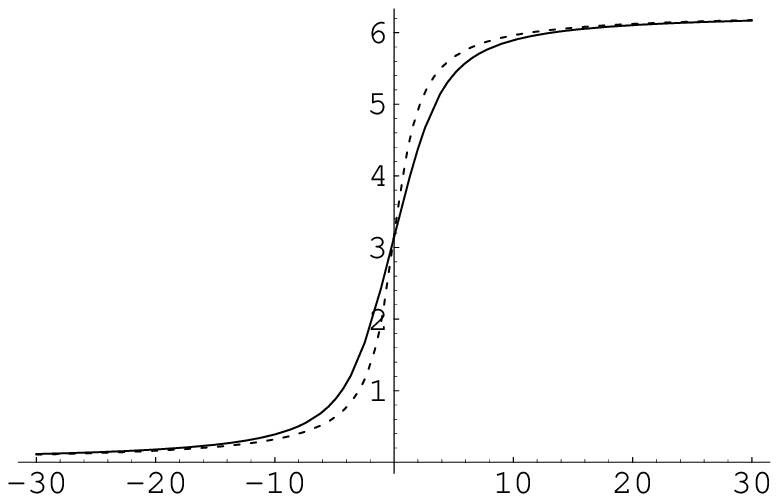}\\
  \caption{Special soliton solutions for Morse intrapair potential (solid lines)
  compared with standard \Y solitons (dotted). Left: the (1,0) soliton; right: the
  (0,1) soliton. See text for parameters values; we plot solutions for $v=0$.}
\end{figure}

\section{Improving the intrapair potential III. \\
Morse potential with a directional term}

As mentioned in the beginning of the previous section, the
Morse-PB potential provides an ``isotropic'' description of the
H-bond mediated intrapair interaction. On the other hand, as
discussed in the Introduction, once we allow rotational motions,
hence the possibility that the alignment between the atoms
intervening in the H-bond is broken, we should consider the
directional nature of the H-bonds. This makes that the bases are
essentially free (as far as this interaction is concerned) once
they are more than a few degrees away from the equilibrium
position.

It should be stressed that in our model the bases cannot come any
nearer than their equilibrium distance; that is, we do not have to
worry about the region on the left of the minimum of $V_p (r)$.

A simple way to modify the Morse-PB potential in order to take
into account the angular dependence of H-bonds is to introduce a
prefactor $A(\vth_1,\vth_2)$ which is unity at equilibrium and
goes rather quickly to zero when at least one of the bases rotate
away from this. It is convenient to choose a smooth function for
this $A$; it should moreover be $2 \pi$-periodic in both its
argument.

The natural candidate for a prefactor with these features is a
gaussian-like function: we choose (no confusion should be possible
with the $\ga$ used in sect.3)
$$ A (\vth_1 , \vth_2 ) \ := \ \exp[- \ga ((1 - \cos \vth_1) + (1 -
\cos \vth_2 ))] \ . \eqno(5.1) $$

We thus have a modified Morse-PB potential
$$ V \ = \ A (\vth_1 , \vth_2 ) \ \[ \a \( (e^{- k \rho (\vth_1,\vth_2) }
- 1)^2 -1 \) \] \eqno(5.2) $$ where we have used notation
introduced in sect.4 above. Adding a constant $\a$ so that $V
(0,0) = 0$, the explicit expression for this is
$$ V \ = \ \a \ \exp[- \ga (1 - \cos \vth_1) (1 -
\cos \vth_2 )] \ \[ \(1 - \exp[- k \rho (\vth_1,\vth_2) ] - 1 \)^2
- 1 \] + \a  \ . \eqno(5.3) $$

We will take the same values as in the previous sections for the
parameters of the Morse potential. As for the newly introduced
parameter $\ga$, this should be chosen so that the H-bond
interaction is conveniently reduced when the relevant atoms are
not properly aligned. A natural criterion is to require $ A
(\theta_0 , \theta_0 ) = a_0 $ with somewhat arbitrary values for
$\theta_0$ and $a_0$. If we choose $\theta_0 = \pi/4$ and $a_0 =
1/10$, we get
$$ \ga \ \simeq \ 4 \ ; \eqno(5.4) $$
this will be our choice. The resulting potential is plotted in
fig.7.

\begin{figure}
  \includegraphics[width=150pt]{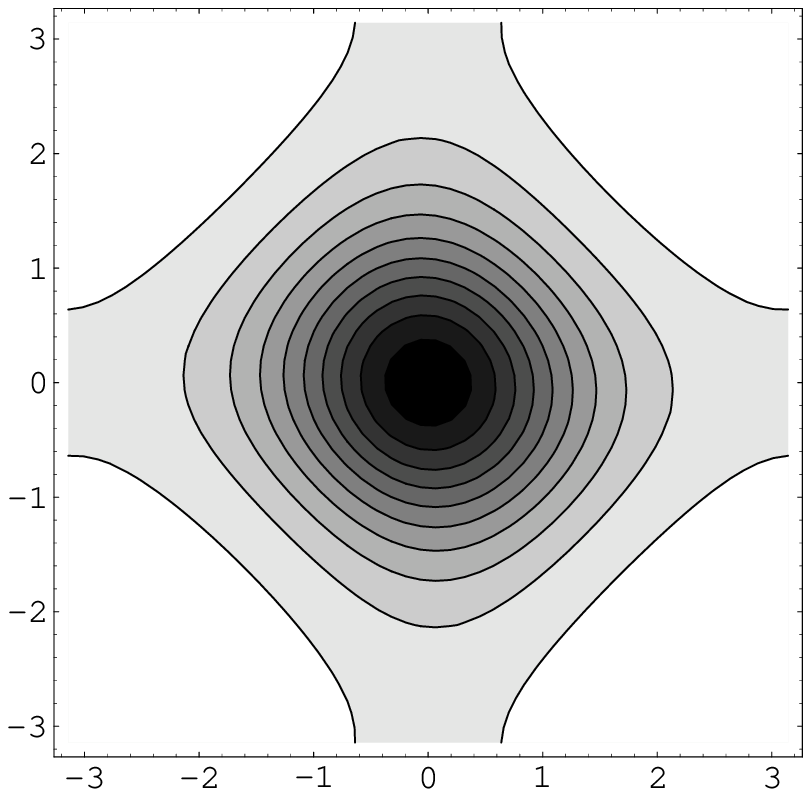}\
  \includegraphics[width=150pt]{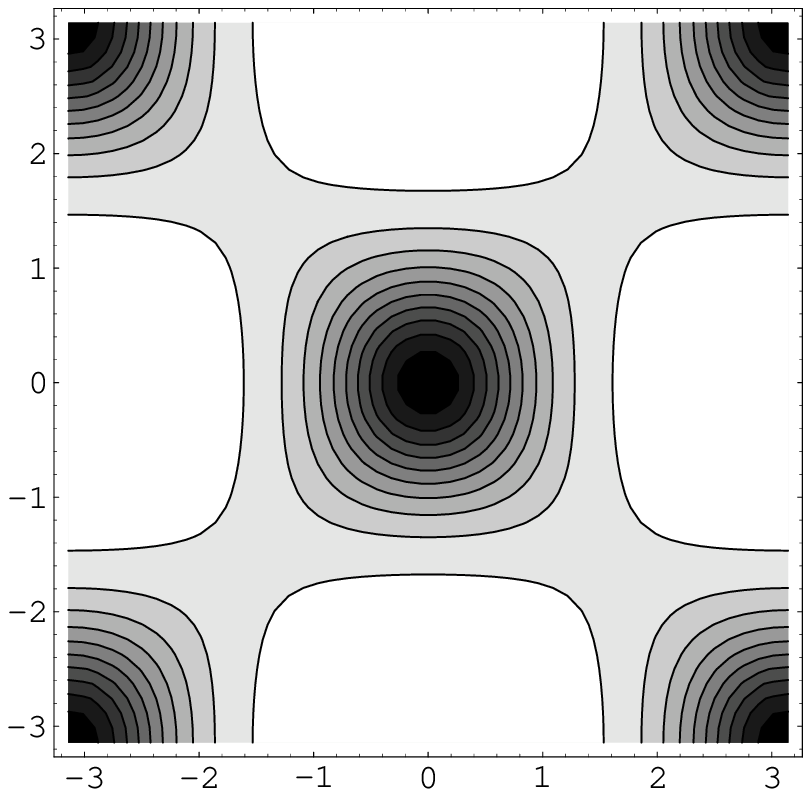}\\
     \caption{The Morse directional potential $V(\vth_1,\vth_2)$ as given
     by (5.3) with $\ga=4$ (left); and the same in adapted coordinates $\psi,\chi$,
     i.e. $V_a (\psi,\chi) = V (\psi+\chi,\psi-\chi)$ (right). These can be
     compared with fig.2.}
\end{figure}

Special soliton solutions are obtained as usual by numerically
solving equations (2.34) and (2.35). The relevant reduced
effective potentials are in this case
$$ \begin{array}{rl}
P (\vphi) \ =& \a \[ 1 \ + \ e^{- 2 \ga (1 - \cos \vphi)} \, \( (
\exp [ - 2 k A F ] - 1 )^2 \, - \, 1 \) \] \ , \\
Q (\eta)  \ =& \a \[ 1 \ + \ e^{- 2 \ga (1 - \cos \eta)} \, \( (
\exp [ - 2 k A G ] - 1 )^2 \, - \, 1 \) \] \ , \end{array}
\eqno(5.5') $$ where we have defined, for ease of writing,
$$
F \, = \, \sqrt{1 - 2 \la \cos \vphi + \la^2} \, - \, (1-\la) \ ,
\ G \, = \, \la \, (1 -  \cos \eta) \ . \eqno(5.5'') $$

\begin{figure}
  \includegraphics[width=150pt]{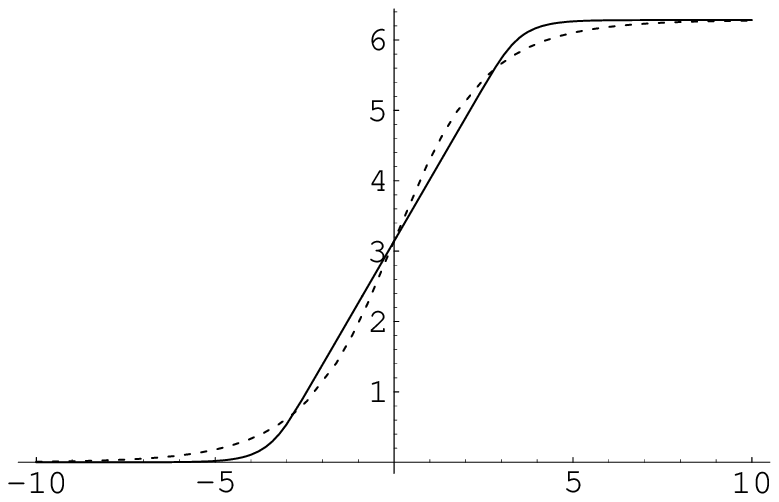}\
  \includegraphics[width=150pt]{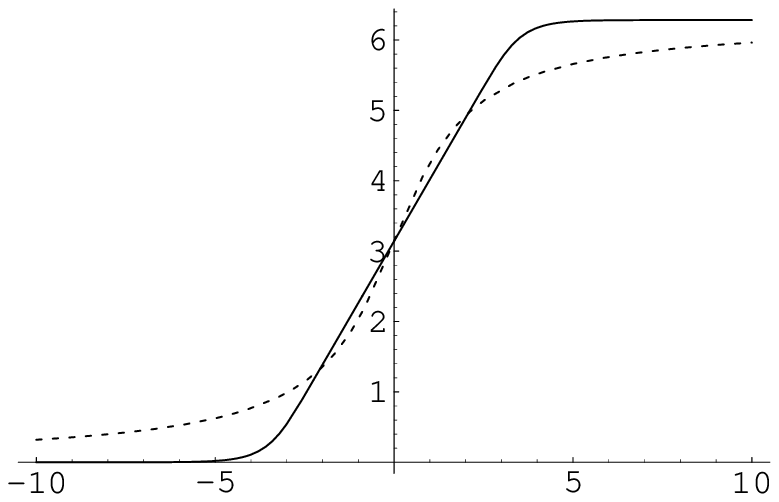}\\
  \caption{The (1,0) soliton (left) and the (0,1) soliton (right) solutions
  for the Morse intrapair potential with directional term (5.3) (solid lines)
  and, for comparison purposes, for the standard \Y model. Here parameters of
  the Morse potential and geometrical parameters are as in section 4, i.e.
  $A = 5.5$ \AA, $r = 4.0$ \AA, $\la = 0.73$, $\a = 0.1$ eV,
  $k = 0.088$ ${\rm \AA}^{-1}$, and $\ga = 4$.}
\end{figure}

Needless to say, our choice for the parameter $\ga$ was to a large
extent arbitrary; we should then check that the resulting soliton
solutions are not strongly dependent on such arbitrary choice. To
this purpose, we have integrated the equations also using
different values for $\ga$, and it resulted that the soliton
solutions are very little dependent on this. The results of
integration with $\ga = 1$ and for $\ga = 10$ are shown, together
with the solution obtained for $\ga = 4$, in fig.9.

\begin{figure}
  \includegraphics[width=150pt]{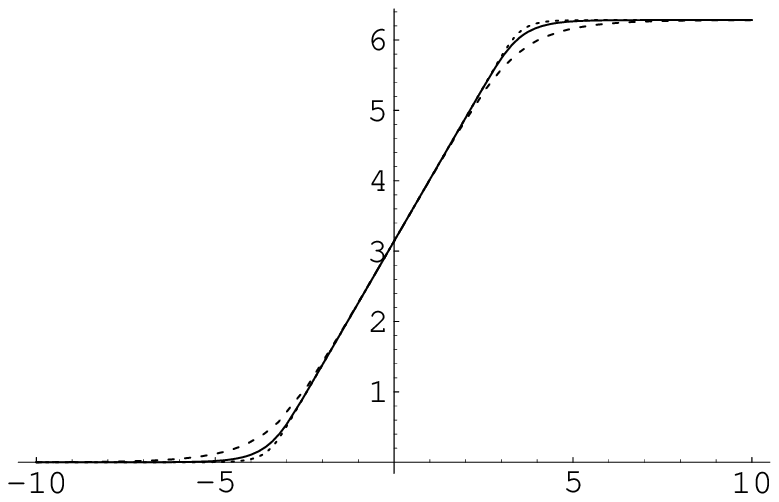}\
    \includegraphics[width=150pt]{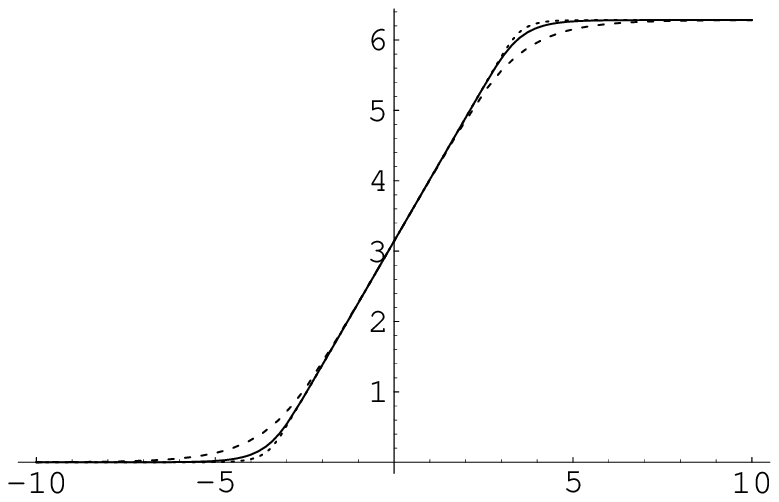}\\
  \caption{Dependence of the (1,0) (left) and (0,1) (right) soliton
  solutions for the Morse intrapair potential with directional term
  on the value of the parameter $\ga$. Keeping all other parameters
  equal (values as in fig.8 above), we plot solutions for
  $\ga=4$ (solid line) together with those for $\ga = 1$ (dashed)
  and for $\ga = 10$ (dotted). Solutions are extremely similar.}
\end{figure}

\section{Discreteness effects}

We have described solitons in the continuum approximation; our
original model was however a discrete one and as well known, going
back to a discretized description the soliton propagation results
to be hindered by the Peierls-Nabarro barrier \cite{PeyNLN}.

The intrapair potentials considered in this note lead to effective
potentials which are steeper than the one corresponding to the
original \Y potential near the equilibrium position, but have a
nearly flat region at larger amplitude of the $\th_i$ fields;
moreover, they lead to soliton solutions which have a much larger
curvature at the interface between the nearly quiescent region and
the ``transition'' region (in which the field passes from 0 to $2
\pi$), and a nearly zero curvature in the transition region. These
effects lead to opposite variation, and so the net effect should
be evaluated quantitatively in order to understand which
contribution is larger.

We thus want to evaluate the Peierls barrier for the different
potentials we have considered; we will focus on the (1,0) special
soliton solutions. The Peierls barrier for a given travelling wave
can be evaluated as follows.

First of all, we can just consider static solutions $\phi (z) =
\vphi(x)$, i.e. set $v=0$; for these the energy of the
configuration will not include the kinetic term, and reduce (see
sect.1) to
$$ H \ = \ U_s \ + U_p \ . \eqno(6.1) $$
Note that the balance between these two contribution is only
marginally dependent on the model we consider: indeed parameters
were determined by the strength of the stacking and the pairing
interactions, as they result from experiments.

The field $\vphi (x)$ should now be replaced by the array of
values
$$ \phi_n \ := \ \vphi (n \de) \ . \eqno(6.2) $$

The stacking energy $U_s$ is given by (1.2). As in our case
$\vth_1 (x) = \vth_2 (x) = \vphi (x)$, the latter reads simply
$$ U_s \ = \ K_s \ \sum_n \, (\phi_{n+1} - \phi_n )^2 \ ; \eqno(6.3)
$$ note this holds for any Y-like model, and in particular
for all the models we have considered.

The pairing energy $U_p$ depends on the intrapair potential $V$,
and is given, see again sect.1 and in particular (1.4), by the sum
of the pairing energy of all base pairs. In our case,
$$ U_p \ = \ \sum_n \, V(\phi_n , \phi_n ) \ = \ \sum_n \,
P(\phi_n) \ . \eqno(6.4) $$

Using (6.2), we can express the energy for the chain configuration
in terms of the soliton field configuration; indeed (6.3) and
(6.4) are also written as
$$
U_s \, = \, K_s \, \sum_n \, [\vphi ((n+1)\de) - \vphi(n \de) ]^2
\ , \ U_p \, = \, \sum_n \, P[\vphi(n \de)] \ . \eqno(6.5) $$

We introduce now a shift parameter $s\in [0,1]$, and compute the
energy when the configuration is shifted by $s \de$:
$$ H(s) \ := \ K_s \ \sum_n \, [\vphi ((n+1+s)\de) - \vphi((n+s) \de) ]^2 \
+ \ \sum_n \, P[\vphi((n + s) \de)] \ . \eqno(6.6) $$ The Peierls
barrier is then given by the different between the maximum and the
minimum over $s\in[0,1]$ of the function
$$ B(s) \ := \ H(s) - H(0) \ . \eqno(6.7) $$

We have computed this for the different models considered here;
the results are summarized in the table below:

\bigskip
\begin{tabular}{|l|r|}
  \hline
  Intrapair potential & Peierls barrier \\
  \hline
  \hline
  Standard Y & $4.3 \cdot 10^{-5}$ eV \\
  Dipole & $2.3 \cdot 10^{-2}$ eV \\
  Morse-PB & $3.2 \cdot 10^{-7}$ eV \\
  Directional Morse ($\ga=4$) & $1.1 \cdot 10^{-2}$ eV \\
  Directional Morse ($\ga=1$) & $3.5 \cdot 10^{-4}$ eV \\
  \hline
\end{tabular}
\bigskip

It results that considering directional rather than isotropic
potentials (that is, dipole-dipole rather than the standard Y
potential; or directional Morse rather than isotropic Morse-PB)
leads to a higher Peierls barrier, i.e. isotropic potentials make
us underestimate its height.

On the other hand, a Morse-type potential (Morse-PB rather than
standard Y; or directional Morse rather than dipole-dipole) lead
to a lower Peierls barrier.

\section{Discussion and conclusions}

We have considered the \Y model for DNA torsion dynamics, in a
version modified by introducing different expressions for the
intrapair potential $V(\vth_1,\vth_2)$; these were:
\par\noindent
(a) a dipole-dipole interaction, which takes into account the
essentially dipolar nature of the H bonds mediating the intrapair
interaction between bases in a pair;
\par\noindent
(b) a Morse-type potential, as often used to model H-bonds;
\par\noindent
(c) a Morse-type potential modified by introducing a prefactor
which takes into account the directional nature of the essentially
dipolar interaction at the origin of H-bonds.

We have focused on travelling soliton solutions, in particular the
special ones -- with topological indices (1,0) and (0,1)
respectively -- in which only one of the fields $\psi$ and $\chi$
varies from the equilibrium configuration.

We found that these solutions vary quite little when we consider
these different expressions for the potential (on the other hand,
the Peierls barrier is quite different in the different cases
considered). In particular, the width of the soliton is, provided
we choose parameters appearing in the model following the same
physical criteria, of the same order of magnitude.

This should not be too surprising: after all, once we fix boundary
conditions -- i.e. the topological indices -- the solution results
from a balance between the stacking interaction (which would favor
very smooth transitions from one vacuum to the other) and the
on-site intrapair potential (which would favor as little bases as
possible being away from minima of this potential). Thus, the
width of the soliton depends essentially on the ration between
strength of these two interactions.

The \Y model is concerned with DNA torsion dynamics; according to
the pioneering proposal of Englander et al. \cite{Eng},
topological (kink) solitons in this dynamics should somehow
correspond to the ``transcription bubbles'' which are formed when
RNA-Polymerase binds to the DNA double chain; from this point of
view, the main success of the \Y model was to provide the correct
size  for these nonlinear excitations.

We found that the soliton solutions have a different shape in the
standard and in the modified \Y models; but their size is not very
different in the different  models. We have also computed the
Peierls barrier for the different models, finding a quite large
range of variability.

In conclusion, our study suggests that:
\begin{itemize}
\item (A) The standard \Y model captures to a large extent the
essential part of DNA dynamics as it can be described by simple
models with a single degree of freedom per nucleotide.

\item (B) On the other hand, study of the standard \Y model can
have led to an overestimation of the Peierls-Nabarro barrier to be
overcome by the topological solitons to move along the discrete
DNA double chain.
\end{itemize}
\bigskip

Point (B) suggests that solitons with too low speed could actually
be unable to propagate along the chain, and thus set also a
minimum speed -- and not just a maximum one, see sect.2 -- for
such solutions, at variance with the analysis of purely continuous
solitonic excitations in the \Y model \cite{GaeSpeed}.

Point (A) shows that it is pointless to attempt an improvement of
the standard \Y model -- e.g. to reconcile the predictions of the
model with different physical requirements, which leads to
contrasting tuning of the parameter values \cite{YakPRE} -- simply
by improving the expressions of the intrapair (pairing) potential.

Albeit we have not considered this point here, the results of this
study and the qualitative discussion above seem to suggest that
the same applies to some extent for improved expressions of the
stacking potential. In this respect we recall, however, that
recent work by Saccomandi and Sgura \cite{SacSgu} points out new
phenomena appearing as a nonlinear stacking potential is
considered.

Thus the main result of our investigation is that in order to go
over the limitation of the standard \Y model we should not try to
investigate more and more detailed description of the interactions
between different parts of the DNA molecule within the description
provided by the standard \Y model, but rather consider a more
detailed model, maybe with equally rough approximation of the
interactions. In the language of \cite{YakPhD}, we should go to a
higher level in the hierarchy of DNA models.

This will be done in a different note \cite{CDG}, where a Y-type
model with more degrees of freedom will be considered. On the
other hand, the investigation reported in the present paper
suggests that the new model can be tackled, in the first
approximation, considering very simple expressions for interaction
potentials.


\end{document}